\documentclass[12pt]{iopart}
\usepackage{iopams}
  
\usepackage{epsfig}  
\newcommand{\beq}{\begin{equation}}
\newcommand{\eeq}{\end{equation}}
\newcommand{\bea}{\vspace{0.25cm}\begin{eqnarray}}
\newcommand{\eea}{\end{eqnarray}}

\newcommand{\pb}{{{\bf p}}}

\def\lsim{\mathrel{\rlap{\lower4pt\hbox{\hskip1pt$\sim$}}
    \raise1pt\hbox{$<$}}}         
\def\gsim{\mathrel{\rlap{\lower4pt\hbox{\hskip1pt$\sim$}}
    \raise1pt\hbox{$>$}}}         

\begin{document}

\title[]{
Parton energy loss in the mini quark-gluon plasma and 
jet quenching in proton-proton collisions 
}

\author{B.G. Zakharov}

\address{
 L.D.~Landau Institute for Theoretical Physics,
        GSP-1, 117940,\\ Kosygina Str. 2, 117334 Moscow, Russia
}
\ead{bgz@itp.ac.ru}
\begin{abstract}
We evaluate the medium suppression of light hadron
spectra in $pp$ collisions at RHIC and LHC energies in the scenario with
formation of a mini quark-gluon plasma. 
We find a significant suppression effect.
For $p_{T}\sim 10$ GeV we obtained the reduction of the spectra
by $\sim [20-30,25-35,30-40]$\% at $\sqrt{s}=[0.2, 2.76,7]$ TeV.
We also discuss how this phenomenon may change the 
predictions for the nuclear modification
factors for $AA$ and $pA$ collisions.
\end{abstract}


\section{Introduction}
The experiments at RHIC and LHC
have provided clear evidence that in $AA$ collisions the hadroproduction 
goes through the formation of a fireball of hot and 
dense quark-gluon plasma (QGP).
This follows from the observation of strong suppression of
high-$p_{T}$ particle spectra (the so-called jet quenching phenomenon)
and from the results of the hydrodynamic simulations of $AA$ collisions.
In the pQCD paradigm the jet quenching is due to radiative 
\cite{GW,BDMPS,LCPI,BSZ,W1,GLV1,AMY}
and collisional  \cite{Bjorken1} energy loss  in the QGP
which soften the parton$\to$hadron fragmentation in $AA$ collisions
(for recent comprehensive reviews, see
\cite{Arleo,Milhano}).
The suppression of the high-$p_{T}$ particle spectra
in $AA$ collisions is characterized by the nuclear
modification factor $R_{AA}$ defined as the ratio
of the particle spectrum in $AA$ collisions to the binary-scaled
spectrum in  $pp$ collisions \cite{JW}
\beq
  R_{AA} = \frac{d\sigma(AA\to hX)/d\pb_{T}dy}{N_{bin} \, 
  d\sigma(pp\to hX)/d\pb_{T}dy}\,.
\label{eq:10}
\eeq
Presently, in theoretical calculations of the $R_{AA}$ 
for the inclusive cross section 
$d\sigma(pp\to hX)/d\pb_{T}dy$
in the denominator in (\ref{eq:10}) predictions of the pQCD are used.
However, if the QGP is produced
in $pp$ collisions as well, the real inclusive cross section
differs from that calculated in pQCD  by its own medium modification
factor $R_{pp}$, i.e.,
\beq
d\sigma_{}(pp\to hX)/d\pb_{T}dy=
R_{pp} d\sigma_{pert}(pp\to hX)/d\pb_{T}dy\,.
\label{eq:20}
\eeq
In this scenario 
the theoretical quantity which should be compared with
the experimental $R_{AA}$ given by (\ref{eq:10}) can be written as
\beq
R_{AA}=R_{AA}^{st}/R_{pp}\,,
\label{eq:30}
\eeq
where $R_{AA}^{st}$ is the standard  nuclear modification factor 
calculated using the pQCD predictions for the particle spectrum in 
$pp$ collisions.
Of course, the $R_{pp}$ is unobservable directly
because experimentally we do not have the baseline spectrum 
with the final state interactions in the QGP switched off. 
Nevertheless, the presence of the $R_{pp}$ in (\ref{eq:30})
may be important for theoretical predictions
for jet quenching in $AA$ collisions. For example,
for the jet flavor tomography of the QGP \cite{Armesto_HQ,BG,RAA12,RAA13}
due to different suppression of light and heavy flavors in $pp$ collisions.

Presently, it is widely believed that in $pp$ collisions
in the studied energy range a hot QCD matter is not produced
in the typical inelastic minimum bias events due to small energy density. 
But in high multiplicity (HM) $pp$ events the energy density
may be comparable to that in $AA$ collisions
at RHIC and LHC energies. And if the thermalization time, $\tau_{0}$, 
is small enough,
say $\tau_{0}\lsim 0.5$ fm, the mini-QGP with size of $\sim 2-3$ fm should be 
formed quite likely to the large-size plasma in $AA$ collisions.
In recent years the possibility that the mini-QGP
may be created in HM $pp$ collisions has 
attracted increasing interest 
(see, for instance, Refs.
\cite{Bozek_pp,Wied_pp,Werner,Camp1,Gyulassy_pp,glasma_pp,SZ}).
Actually, we already have some experimental indications in favor 
for the formation of the mini-QGP in HM $pp$ collisions.
It is possible that 
the ridge correlation structure in HM $pp$ events at
$\sqrt{s}=7$ TeV observed  by the CMS collaboration  \cite{CMS_ridge1} 
is  due to  the transverse flow of the QGP. 
In \cite{Camp1},
employing Van Hove's idea \cite{VH} that  
phase transition should  lead to anomalous behavior of 
the mean transverse momentum $\langle p_{T}\rangle$ as a function of 
multiplicity, it has been argued that the data on 
$\langle p_{T} \rangle$ 
signal possible plasma formation in the domain $dN_{ch}/d\eta\sim  6-24$.
Some intriguing similarities between the results of the femtoscopic
analyses of $pp$ and $AA$ collisions at RHIC \cite{STAR_fem}
and LHC \cite{ALICE_fem} also may signal the formation of the
collective QCD matter in the HM $pp$ events.
The preliminary data from 
ALICE \cite{ALICE_jet_UE}, indicating
that for the HM $pp$ events jets undergo a softer fragmentation,
also support this idea.

From the point of view of jet quenching it is important that 
the conditions for the QGP production in $pp$ collisions are better
in events with jets, because the multiplicity of soft off-jet particles 
(the so-called underlying events (UE), see \cite{Field} for a review) 
is enhanced by a factor of $2-3$ \cite{CDF} (below we denote this
factor by $K_{ue}$)
as compared to the minimum bias multiplicity. And even at RHIC 
energies $\sqrt{s}\sim 0.2$ TeV 
the UE multiplicity may be high enough for the QGP formation.
In our recent work \cite{Z_pp} we have evaluated the medium modification
of the fragmentation functions (FFs) for $\gamma$-triggered and inclusive 
jets in HM $pp$ collisions, and have presented preliminary results for medium
suppression of hadron spectra.
We have found that the medium effects are surprisingly strong. 
In the present work we perform a detailed analysis
of the medium modification of the hadron spectra in $pp$ collisions 
due to parton energy loss in the mini-QGP.
We evaluate $R_{pp}$ of charged hadrons in the central rapidity
region ($y=0$)
at RHIC ($\sqrt{s}=0.2$ TeV) and LHC ($\sqrt{s}=2.76$ and $7$ TeV) energies.
We also address the effect of $R_{pp}$ on $R_{AA}$
at RHIC and LHC energies
and on $R_{pA}$
in the context of the recent data from ALICE
\cite{ALICE_RpPb} on $R_{pPb}$ at $\sqrt{s}=5.02$ TeV.
The analysis is based on the light-cone path integral (LCPI) approach 
\cite{LCPI,BSZ} to induced gluon emission. It treats accurately 
the finite-size and Coulomb effects (which are very important 
for the mini-QGP), 
the mass effects, and is valid beyond the soft gluon approximation.
We evaluate the medium modified FFs within  
the scheme developed previously for $AA$ collisions \cite{RAA08}.
It takes into account both radiative and collisional energy loss. 
Previously in \cite{RAA11,RAA12,RAA13}  the approach has been 
successfully used for description of jet quenching in $AA$ collisions.

The paper is organized as follows. In the next section we 
discuss the parameters of the mini-QGP for the UE $pp$ events
at RHIC and LHC.
In section 3 we discuss the basic aspects of the theoretical framework.
In section 4 we present the numerical results on parton energy loss
in the mini-QGP and the medium modification factors 
for $pp$, $AA$ and $pA$ collisions,
section 5 summarizes our work.

\section{Mini-QGP in proton-proton collisions}
We neglect the transverse expansion of the mini-QGP and 
use 1+1D Bjorken's model \cite{Bjorken2}, which gives 
$T_{0}^{3}\tau_{0}=T^{3}\tau$. 
For $\tau<\tau_{0}$ we take medium density $\propto \tau$. 
In the basic variant we take $\tau_{0}=0.5$ fm. 
Approximately such $\tau_{0}$ is used in most studies of jet
quenching in $AA$ collisions. For the QGP
produced in $AA$ collisions with the lifetime/size $L\gg \tau_{0}$ 
the medium modification of hadron spectra is not very sensitive to 
variation of $\tau_{0}$. But this may be untrue for the 
mini-QGP in $pp$ collisions when the plasma size is not very large as
compared to $\tau_{0}$. 
To understand the sensitivity
of $R_{pp}$ to $\tau_{0}$, which is not well constrained 
by the hydrodynamic modeling of $AA$ collisions \cite{Heinz_hydro}, 
we also perform calculations for $\tau_{0}=0.8$ fm.
As in our analyses of $AA$ collisions \cite{RAA08,RAA11,RAA12}, we neglect 
variation of the initial temperature $T_{0}$ with the 
transverse coordinates.
To fix $T_{0}$ 
we use the entropy/multiplicity ratio 
$C=dS/dy{\Big/}dN_{ch}/d\eta\approx 7.67$ obtained in \cite{BM-entropy}.
The initial entropy density can be written as 
\beq
s_{0}=\frac{C}{\tau_{0}\pi R_{f}^{2}}\frac{dN_{ch}}{d\eta}\,,
\label{eq:40}
\eeq
where $R_{f}$ is the radius of the created mini-QGP fireball.
We ignore the azimuthal anisotropy, and regard the $R_{f}$ as an effective
plasma radius, which includes $pp$ collisions at all impact parameters.
This approximation seems to be plausible since anyway the jet production 
should be dominated by the almost head-on collisions
for which the azimuthal effects should be small. 
This is supported by calculation of the 
distribution of jet production cross section in the impact parameter plane 
using the MIT bag model
which says that only $25$\% of jets come from
$pp$ collisions with the impact parameter larger than the bag radius.
It says that typically the fireball has a relatively small eccentricity.
Anyway, we are interested in $R_{pp}$, which is averaged over the azimuthal
angle, and it is practically insensitive to the fireball eccentricity.  

One can expect that
for $pp$ collisions the typical radius of the fireball should be
about the proton radius $R_{p}\sim 1$ fm.     
It agrees qualitatively with $R_{f}$ obtained 
for $pp$ collisions at $\sqrt{s}=7$ TeV in numerical simulations
performed in \cite{glasma_pp} 
within the IP-Glasma model \cite{IPG12}. The $R_{f}$ from \cite{glasma_pp}
grows approximately as linear function of $(dN_{g}/dy)^{1/3}$ and then 
flatten. The flat region corresponds to almost head-on
collisions. In this regime the fluctuations of multiplicity are 
dominated by the fluctuations of the glasma color fields \cite{glasma_pp}. 
We use the $R_{f}$ from \cite{glasma_pp} 
parametrized in \cite{RPP}  via $dN_{g}/dy$ in the form
\beq
 R_{f} = 1\,{\rm fm}\times f_{pp}\left(\sqrt[3]{dN_g/dy}\right)
\label{eq:50_1}
\eeq
with
\beq
 f_{pp}(x) = 
\left\{ \begin{array}{ll}
         0.387 + 0.0335 x + 0.274\,x^2 - 0.0542\,x^3  & \mbox{if $x < 3.4$,}\\
         1.538 & \mbox{if $x \geq 3.4 $.}
\end{array} \right.
\label{eq:50} 
\eeq
We evaluate $R_{f}$ taking $dN_{g}/dy=\kappa dN_{ch}/d\eta$
with $\kappa=C45/2\pi^{4}\xi(3)\approx 2.13$.
Possible increase of the $R_{f}$ from RHIC to LHC should not be 
important since our results are not very sensitive to variation of $R_{f}$.

The multiplicity density of the UEs grows with momentum of the 
leading charged jet hadron at $p_{T}\lsim 3-5$ GeV and then flatten 
\cite{CDF,PHENIX_dA,ATLAS_UE_Nch,CMS_UE_Nch,ALICE_UE_Nch} (in terms of the jet
energy the plateau region corresponds approximately to 
$E_{jet}\gsim 15-20$ GeV).
To fix the $dN_{ch}/d\eta$ in (\ref{eq:40}) at $\sqrt{s}=0.2$ TeV 
we use the UE multiplicity enhancement factor $K_{ue}$ 
from PHENIX \cite{PHENIX_dA} obtained by dihadron correlation
method. Taking for the minimum bias 
non-diffractive events $dN_{ch}^{mb}/d\eta=2.98\pm 0.34$ from STAR 
data \cite{STAR-dnch}, we obtained for the UEs in the plateau region 
$dN_{ch}/d\eta\approx 6.5$.  
To evaluated the UE multiplicity  
at $\sqrt{s}=2.76$ and $5.02$ TeV we use the data from ATLAS 
\cite{ATLAS_UE_Nch} at $\sqrt{s}=0.9$ and $7$ TeV that give in the 
plateau region $dN_{ch}/d\eta\approx 7.5$ and $13.9$.
Assuming that $dN_{ch}/d\eta\propto s^{\delta}$ by interpolating
between $\sqrt{s}=0.9$ TeV and 7 TeV we obtained for  
the UE multiplicity density in the plateau region $dN_{ch}/d\eta\approx 10.5$
and $12.6$ at $\sqrt{s}=2.76$ and $5.02$ TeV, respectively.
With the above values of the UE multiplicity densities in the plateau regions
we obtain the following 
values for the fireball radii
\beq
R_{f}[\sqrt{s}=0.2,2.76,5.02,7\,\, \mbox{TeV}]
\approx[1.3,1.44,1.49,1.51]\,\,\mbox{fm}\,.
\label{eq:60}
\eeq
With these radii, using (\ref{eq:40}) and the ideal gas 
formula $s=(32/45+7N_{f}/15)T^{3}$ (with $N_{f}=2.5$),
we obtain the initial temperatures of the QGP
\beq
T_{0}[\sqrt{s}=0.2,2.76,5.02,7\,\,\mbox{TeV}]
\approx[199,217,226,232]\,\,\mbox{MeV}\,.
\label{eq:70}
\eeq
One can see that the values of $T_{0}$ lie well above 
the deconfinement temperature $T_{c}\approx 160-170$ MeV 
\cite{T_c1,T_c2}\footnote{In fact, if one uses the entropy from
the lattice calculations \cite{T_c1,T_c2} the fireball temperatures
in (\ref{eq:70}) will be higher by $\sim 10-15$\%. We ignore
this difference since for jet quenching the crucial quantity
is the entropy, which we take from experimental data (see discussion
below in Sec. IV).}. 
For such initial temperatures the purely plasma phase may exist up to 
$\tau_{QGP}\sim 1-1.5$ fm, and beyond $\tau_{QGP}$ the hot QCD matter 
will evolve in the mixed phase up to $\tau_{max}\sim 2R_{f}$ where the transverse
expansion should lead to a fast cooling of the system. Since in the
interval $\tau_{QGP}< \tau < \tau_{max}$ the QGP fraction 
in the mixed phase is approximately $\propto 1/\tau$ \cite{Bjorken2} we can use
in calculating  jet quenching the $1/\tau$ dependence of the number
density of the scattering centers in the whole range of $\tau$
(but with the Debye mass defined for $T\approx T_{c}$ at $\tau> \tau_{QGP}$).

Although we neglect the transverse expansion of the QCD
matter, it should not lead to large errors in our predictions.
As was demonstrated in \cite{BMS_hydro} the transverse motion
does not affect strongly jet quenching in $AA$ collisions.
Physically it is due to an almost complete compensation between the enhancement 
of the energy loss caused by increase of the medium size and its 
suppression caused by reduction of the medium density.
In $pp$ collisions the effect should be even smaller 
since the typical formation length for induced gluon emission is of 
the order of $R_{f}$ or larger. In such a regime 
the parton energy loss is mostly controlled by 
the mean amount of the matter traversed by fast partons, and the details of 
the density profile along the jet trajectory are not very important.
Also, 
in $pp$ collisions the QCD matter 
spends much time in the mixed phase, where the sound velocity 
becomes small and the transverse expansion should be less intensive
than in $AA$ collisions.

We conclude this section with two additional remarks. First,
naively one could think that for evaluation of the
medium suppression of the minimum bias $pp$ spectrum one should
use the minimum bias multiplicity density in evaluating the mini-QGP
parameters. But it would be wrong. Indeed, the minimum bias events
include events with and without jet production, and the minimum
bias high-$p_{T}$ spectrum is related to events with jet (at least one)
production. The corresponding multiplicity density for such events
is exactly the UE $dN_{ch}/d\eta$. 

The second remark concerns the formula
(\ref{eq:40}). It implicitly assumes that the UE multiplicity distribution
$dN_{ch}/d\eta$, likewise the minimum bias multiplicity density,
has a central plateau in rapidity (we assume that jet is produced at 
$y=0$). For typical inelastic events the existence of the central
plateau is a consequence of the approximate longitudinal boost invariance.
In the glasma picture \cite{glasma_pp} it naturally appears due
to boost invariance of the initial glasma color fields. 
However, for the UE events this invariance is broken by the jet production 
at $y=0$. And in principle there may be a bump in the UE multiplicity
distribution near the jet rapidity, say due to the initial state 
radiation. For (\ref{eq:40}) to be applicable the half-width of the bump should 
satisfy the inequality 
\beq
\Delta \eta\gsim c_{s}\ln(\tau_{f}/\tau_{0})\,,
\label{eq:79}
\eeq
where $\tau_{f}$ is the freezeout time and $c_{s}$ is the sound velocity of 
the matter. 
This inequality ensures that
in the whole interval from $\tau_{0}$ to $\tau_{f}$ the edges of the 
bump do not affect the $\tau$-dependence of the entropy $s\propto 1/\tau$.
For the relevant temperature
range $c_{s}\lsim 0.5$ \cite{T_c2}.
Then taking $\tau_{f}\sim \tau_{max}$ and  $\tau_{max}/\tau_{0}\sim 6$ 
we obtain from (\ref{eq:79}) 
$\Delta \eta\gsim 1$. The fact that data on the UE $dN_{ch}/d\eta$ 
from ATLAS \cite{ATLAS_UE_Nch} obtained for $|\eta|<2.5$ and from ALICE
\cite{ALICE_UE_Nch} obtained for $|\eta|<0.8$ agree between each other 
says that the above inequality is satisfied. And the formula (\ref{eq:40})
can be safely used for the UEs.

\section{Sketch of the Calculations}
We now turn to the jet quenching in the mini-QGP produced 
in $pp$ collisions.
Our treatment 
is similar to that for $AA$ collisions in 
our previous analysis \cite{RAA08} to which we 
refer the reader for details.
Here we give only a brief sketch of the calculations,
focusing on aspects relevant for jet quenching in
the mini-QGP, and present parameters of the model.
\subsection{Perturbative and medium modified inclusive cross sections}
As usual we write the perturbative inclusive cross section in (\ref{eq:20}) 
in terms of the vacuum parton$\to$hadron FF $D_{h/i}$
\beq
\frac{d\sigma_{pert}(pp\to hX)
}{d\pb_{T} dy}=
\sum_{i}\int_{0}^{1} \frac{dz}{z^{2}}
D_{h/i}^{}(z, Q)
\frac{d\sigma(pp\to iX)}{d\pb_{T}^{i} dy}\,,
\label{eq:80}
\eeq
where
${d\sigma(pp\to iX)}/{d\pb_{T}^{i} dy}$ 
is the ordinary hard cross section,
$\pb_{T}^{i}=\pb_{T}/z$ is the parton 
transverse momentum. 
We write the real inclusive cross section, which accounts for the 
final state interactions in the QCD matter,
in a similar form but with the medium modified FF
$D_{h/i}^{m}$
\beq
\frac{d\sigma_{}(pp\rightarrow hX)}{d\pb_{T} dy}=
\sum_{i}\int_{0}^{1} \frac{dz}{z^{2}}
D_{h/i}^{m}(z, Q)
\frac{d\sigma(pp\rightarrow iX)}{d\pb_{T}^{i} dy}\,.
\label{eq:90}
\eeq
Here it is implicit that $D_{h/i}^{m}$ 
is averaged over the geometrical variables of the hard parton process and 
the impact parameter of  $pp$ collision.

The formula (\ref{eq:90}) can be viewed as an analogue of the formula
for the minimum bias $R_{AA}$ defined in the whole centrality 
(impact parameter) range. 
However, there is one important difference between $pp$ and $AA$ collisions. 
In $AA$ collisions at a given impact parameter the fluctuations of the
multiplicity (and of the parameters of the fireball) are small. 
And this allows to relate the centrality (defined through the multiplicity)
to the impact parameter. 
In $pp$ collisions one cannot relate the multiplicity density to the
impact parameter, since 
for each impact parameter and jet production point in the transverse
plane (which can be localized with the accuracy $\sim z/p_{T}$)
the fluctuations of the multiplicity density are large.
These fluctuations, together with the event-by-event fluctuations 
of the impact parameter and the jet production point, give 
the observable fluctuating UE $dN_{ch}/d\eta$, which can be translated into
the fluctuating fireball parameters.
However, the detailed dynamics of the UEs and  of the multiplicity fluctuations
in such events is far from being clear. In particular, we do not know
whether the enhancement of the UE multiplicity is only due to
the fact that the jet production is biased to more central collisions
and to which extent it may be related to the increase of the soft gluon
density in jet production due to the initial state radiation.
Therefore an accurate accounting for the fluctuations of the parameters
of the mini-QGP fireball is impossible.
In the present study 
in evaluating $D_{h/i}^{m}$ we take  into account (approximately)  only the 
event-by-event variations of the geometrical parameters (see below), 
but ignore the fluctuations of the UE $dN_{ch}/d\eta$.
And evaluate the parameters of the fireball simply using 
the typical UE multiplicity density,
although technically the inclusion of the fluctuations of 
the UE $dN_{ch}/d\eta$ in our formalism is quite simple, 
and we do it to estimated the accuracy of our approximation (see below). 


As in \cite{RAA08}, we calculated the hard cross sections 
in  the LO pQCD with the CTEQ6 \cite{CTEQ6} parton distribution functions 
(PDFs).
To simulate the higher order effects in  calculating the partonic
cross sections we take for the virtuality scale in $\alpha_{s}$ the value 
$cQ$ with $c=0.265$ as in the PYTHIA event generator \cite{PYTHIA}.
This gives a fairly good description of the $p_{T}$-dependence of the 
spectra in $pp$ collisions. Of course, in principle, in the scenario 
with the QGP formation for a fully consistent treatment of $R_{pp}$
(and $R_{AA}$) one should use a bootstrap procedure
and compare with the experimental
data not the perturbative cross section (\ref{eq:80}) but 
the real one given by (\ref{eq:90}), and namely the latter should be adjusted 
(say, by varying PDFs, FFs, and $\alpha_s$)
to describe experimental data. However, since the hadron spectra 
have very steep $p_{T}$-dependence (as compared to 
a relatively weak $p_{T}$-dependence of $R_{pp}$)
this inconsistency may be safely ignored in calculating 
$R_{pp}$ (the same is true for $R_{AA}$ and $R_{pA}$).

For the hard scale $Q$ in the FFs in (\ref{eq:80}), 
(\ref{eq:90}) we use $p_{T}/z$.
We calculate the vacuum FFs $D_{h/j}$  
as a convolution of the KKP \cite{KKP} parton$\to$hadron FFs at 
soft scale $Q_{0}=2$ GeV with the DGLAP parton$\to$parton FFs 
$D_{j/i}^{DGLAP}$
describing the evolution from $Q$ to $Q_{0}$. The latter have been 
computed with the help of PYTHIA \cite{PYTHIA}. 
This procedure reproduces well the whole $Q$-dependence of the KKP 
\cite{KKP} parametrization of the vacuum FFs. 
For a given fast parton path length in the QGP 
the medium modified FFs $D_{j/i}^{m}$ have been calculated in a similar way but
inserting between the DGLAP parton$\to$parton FFs and the KKP
parton$\to$hadron
FFs the parton$\to$parton FFs $D_{j/i}^{ind}$ which correspond 
to the induced radiation stage in the QGP.
The induced radiation FFs $D_{j/i}^{ind}$ have been calculated
from the medium induced gluon spectrum using Landau's 
method \cite{BDMS_RAA} imposing the flavor and momentum conservation
(again, we refer the interested reader to \cite{RAA08} for details).

Note that, since in both the vacuum and the medium modified FFs
the DGLAP evolution is accounted for in the same way,
the medium effects vanish strictly at zero matter density, as it must be.
The above  approximation with the time ordered and independent
DGLAP and induced radiation stages,
suggested for the large-size plasma
produced in $AA$ collisions \cite{RAA08}, seems to be reasonable 
for the mini-QGP as well
(at least in the jet energy region $\lsim 30-50$ GeV 
where the suppression effect appears to be strongest)
since the typical formation time for the most energetic DGLAP gluons 
is of the order of (or smaller) than the thermalization time $\tau_{0}$.
It is worth noting that, although the time ordering of the DGLAP and 
induced radiation stages seems to be physically reasonable, 
the permutation of these stages in the above convolution gives a
very small effect \cite{RAA08}.

Since we do not consider the azimuthal effects, the averaging of 
the medium modified FFs over the geometrical
variables of the hard parton process and $pp$ collisions 
in the impact parameter plane is simply reduced to
averaging over the parton path length $L$ in the QGP.
It cannot be performed accurately since 
the distribution of hard processes in the impact parameter 
plane  is not known yet. But one can expect that
the effect of $L$ fluctuations should be relatively
small for any more or less centered distribution
of energetic partons in the proton wave function.
We have performed averaging over $L$ 
using the distribution of hard processes in the impact parameter plane 
obtained with the quark distribution from the MIT bag model
(we assume that the valence quarks and the hard gluons 
radiated by the valence quarks follow approximately the same 
distribution in the transverse spacial coordinates).
Calculations within this model show that
practically in the full range of the impact parameter of 
$pp$ collisions the distribution
in $L$ is sharply peaked around $L\approx\sqrt{S_{ov}/\pi}$, where $S_{ov}$
is the overlap area for two colliding bags. It means that our 
fireball radius $R_{f}$ (which includes all centralities) at the same
time gives the typical path length for fast partons.
Our calculations show that the effect of the $L$-fluctuations
on $R_{pp}$ is relatively small. As compared to $L=R_{f}$ they
reduce the medium modification by $\sim 10-15$\%.

As in our previous studies of jet quenching in $AA$ collisions 
we treat the collisional energy loss, which is relatively 
small \cite{Z_Ecoll}, as a small perturbation to 
the radiative mechanism. 
We incorporate it in the above procedure simply by renormalizing 
the QGP temperature in calculating the medium modified FFs
(see \cite{RAA08} for details). We assume that the collisional energy loss
vanishes at $\tau<\tau_{0}$ in the pre-equilibrium stage which probably
is populated by strong collective glasma color fields, and the concept
of the collisional energy loss is hardly applicable in this region.
On the contrary, it is clear that the coherent glasma fields
can give some contribution to the radiative energy loss
(probably rather small \cite{AZ_glasma}).
For this reason the use of the linearly growing
plasma density at $\tau<\tau_{0}$ seems to be a 
plausible parametrization to model the transition
from the glasma phase to the hydrodynamically evolving 
QGP, which of course cannot be abrupt.

\subsection{Medium induced gluon spectrum and parameters of the model}
As in \cite{RAA08} we evaluate the medium induced gluon spectrum
$dP/dx$ ($x=\omega/E$ is the gluon fractional momentum)
for the QGP modeled by a 
system of the static Debye  screened color centers \cite{GW}. 
We use the Debye mass obtained in the lattice calculations \cite{Bielefeld_Md} 
giving $\mu_{D}/T$ slowly decreasing with $T$  
($\mu_{D}/T\approx 3.2$ at $T\sim T_{c}$, $\mu_{D}/T\approx 2.4$ at 
$T\sim 4T_{c}$). 
For the quasiparticle masses of light quarks and gluon
in the QGP we take $m_{q}=300$ and $m_{g}=400$ MeV  supported by 
the analysis of the lattice data \cite{LH}.
But the results are
not very sensitive to the $m_{g}$, and practically
insensitive to the value of $m_{q}$.
We evaluated the induced gluon spectrum using the 
representation suggested in \cite{Z04_RAA}. 
It expresses the $x$-spectrum 
for gluon emission from a quark (or gluon) 
through the light-cone wave function 
of the $gq\bar{q}$ (or $ggg$) system in the coordinate $\rho$-representation.
The $z$-dependence of the wave function is governed by
a two-dimensional Schr\"odinger equation 
with the ``mass'' $\mu=x(1-x)E$ ($E$ is the initial parton energy)
in which the
longitudinal coordinate $z$ 
plays the role of time and the potential $v(\rho)$ is proportional 
to the local plasma density/entropy times a linear combination of
the dipole cross sections $\sigma(\rho)$, $\sigma((1-x)\rho)$
and $\sigma(x\rho)$.
Note that the physical pattern of induced gluon emission in the mini-QGP
differs from that for the large-size QGP.
For the mini-QGP when the typical path length in the medium 
$L\sim 1-1.5$ fm the energy loss is dominated by gluons with
$L_{f}\gsim L$, where $L_{f}\sim 2\omega/m_{g}^{2}$ is the gluon formation
length in the low density limit. 
It is the diffusion regime in the terminology of \cite{Z_OA},
in which the finite-size effects play a crucial role.
In this regime the dominating contribution comes from the $N=1$ rescattering
and the Coulomb effects are very important \cite{Z_OA}. 
On the contrary, for the QGP in $AA$ collisions a considerable part
of the induced energy loss comes from gluons with $L_{f}\lsim L$.
Indeed, in the bulk of the large-size QGP $L_{f}\sim 2\omega S_{LPM}/m_{g}^{2}$,
where $S_{LPM}$ is the LPM suppression factor.
For RHIC and LHC typically $S_{LPM}\sim 0.3-0.5$ for $\omega\sim 2$ GeV, it 
gives $L_{f}\sim 1.5-2.5$ fm which is smaller than the typical $L$
for the QGP in $AA$ collisions.
In this regime the finite-size effects are much less important
and induced gluon radiation is (locally) approximately similar to that 
in an infinite extent matter.

From the point of view of jet quenching in $pp$ collisions 
it is important that induced radiation in the mini-QGP
is more perturbative than in the QGP in $AA$ collisions.
Indeed, let us consider induced radiation for the mini-QGP. 
From the Schr\"odinger diffusion relation one can obtain for 
the typical transverse size of the three parton system 
$\rho^{2}\sim 2\xi/\omega$, where 
$\xi$ is the path length after gluon emission.
Then, using the fact that $\sigma(\rho)$ is dominated by the $t$-channel
gluon exchanges with virtualities up to $Q^{2}\sim 10/\rho^{2}$  
\cite{NZ_piki} we obtain $Q^{2}\sim 5\omega/\xi$. For $\omega\sim 2$
and $\xi\sim 0.5-1$ fm it gives rather large virtuality scale $Q^{2}\sim 2-4$
GeV$^{2}$. The virtuality scale for $\alpha_{s}$ in the gluon emission
vertex has a similar form but smaller by a factor of $\sim 2.5$ \cite{Z_Ecoll}.
The $1/\xi$ dependence of $Q^{2}$ persists up to $\xi\sim L_{f}$.
For the large-size QGP in the above formulas one should replace $\xi$
by the real in-medium $L_{f}$ (which contains $S_{LPM}$) which
is by a factor of $\sim 2$ larger than the typical values of $\xi$ 
for the mini-QGP. It results in a factor of $\sim 2$
smaller virtualities for the QGP in $AA$ collisions. 
In this sense the calculations for the mini-QGP
are more robust than for the large-size QGP.

As in \cite{RAA08,RAA11,RAA12,RAA13} 
we perform calculations 
of radiative and collisional energy 
loss with running $\alpha_{s}$
frozen at some value $\alpha_{s}^{fr}$ at low momenta.
For gluon emission in vacuum a reasonable choice is 
$\alpha_{s}^{fr}\sim  0.7-0.8$ \cite{NZ_HERA,DKT}. But 
in plasma thermal effects can suppress $\alpha_{s}^{fr}$.
However, in principle, the extrapolation from the vacuum gluon 
emission  to the induced radiation is unreliable due to large 
theoretical uncertainties of jet quenching calculations. For this
reason $\alpha_{s}^{fr}$ should be treated as a free parameter
of the model. To evaluate the medium suppression in $pp$ collisions
it is reasonable to use the information on the values of $\alpha_{s}^{fr}$
necessary for description of jet quenching in $AA$ collisions.
Previously we have observed \cite{RAA13} that data on $R_{AA}$ 
are consistent with $\alpha_{s}^{fr}\sim 0.5$ for RHIC
and $\alpha_{s}^{fr}\sim 0.4$ for LHC.
The reduction of $\alpha_{s}^{fr}$ from RHIC to LHC 
may be related to stronger thermal effects in the QGP due to 
higher initial temperature at LHC. 
But the analysis \cite{RAA13} is performed under assumption that
there is no medium suppression in $pp$ collisions.
The inclusion of $R_{pp}$ should increase the values of $\alpha_{s}^{fr}$.
However, in \cite{RAA13} we used the plasma density 
vanishing at $\tau<\tau_{0}$, whereas in the present work we use 
the QGP density $\propto \tau$ in this region which leads
to stronger medium suppression.
As a result, as we will see 
below, the values of $\alpha_{s}^{fr}$, which are preferable
from the standpoint of the description of the data on $R_{AA}$,
remain approximately the same, or a bit larger, as obtained 
in \cite{RAA13}.
If the difference between the preferable values of 
$\alpha_{s}^{fr}$  for $AA$ collisions at RHIC and LHC
is really due to the thermal effects, 
then  for the mini-QGP with $T_{0}$ as given in 
(\ref{eq:70}) a reasonable window  is 
$\alpha_{s}^{fr}\sim 0.6-0.7$. 
In principle for the mini-QGP the thermal reduction 
of $\alpha_{s}$ may be smaller than that for the large-size 
plasma (at the same temperature). 
Since at $L_{f}\lsim L$, which typically holds for the mini-QGP,
the dominating contribution to the induced gluon spectrum
comes from configurations with interference of the emission amplitude 
and complex conjugate one when one of them has the gluon emission 
vertex outside the medium and is not affected by the medium effects at all.
We perform the calculations for
$\alpha_{s}^{fr}=0.5$, $0.6$ and $0.7$.
Note that, in principle $R_{pp}$ should be less sensitive to
$\alpha_{s}^{fr}$ than $R_{AA}$ since, as we said above, the typical
virtualities for induced gluon emission in the mini-QGP are
larger than that in the large-size QGP.
As will be seen from our numerical results, the sensitivity to
$\alpha_{s}^{fr}$ is really quite weak.

\section{Numerical Results}
\subsection{Energy loss in the mini-QGP}
Before presenting the results for the medium modification factors 
it is worthwhile first to show the results for radiative and 
collisional energy loss that may give some insight into the magnitude
of the medium effects generated by the mini-QGP in $pp$ collisions.
In Fig.~1 we show the energy dependence of the total 
(radiative plus collisional) and collisional
energy loss for partons produced in the center of the mini-QGP 
fireball for RHIC and LHC conditions for $\alpha_{s}^{fr}=0.6$.
Both the radiative and collisional contributions are 
calculated for the lost energy smaller than half of the initial parton
energy. 
The fireball radius $R_{f}$ and the initial temperature $T_{0}$ have
been calculated with the UE multiplicity density dependent on the
jet energy $E$ using the data \cite{PHENIX_dA,ATLAS_UE_Nch}. 
In \cite{PHENIX_dA,ATLAS_UE_Nch} the UE activity has been measured 
vs the transverse momentum of the leading charged jet hadron
(we denote it as $p_{T}^{l}$).
To obtain the UE $dN_{ch}/d\eta$ as a function of the jet energy $E$
we neglect the fluctuations of $p_{T}^{l}$ for a given $E$
and use the rigid relation  $p_{T}^{l}=\langle z_{l}\rangle E$,
where $\langle z_{l}\rangle$ is the average fractional momentum of the leading
jet hadron. For the $\langle z_{l}\rangle$ we take 
the PYTHIA predictions, which gives in the relevant energy region 
($E\lsim 10$ GeV) $\langle z_{l}\rangle \sim 0.26$ for gluon jets
and 
$\langle z_{l}\rangle \sim  0.36$ for quark jets.
The jet energy dependence of the parameters of the fireball becomes
important only for partons with $E\lsim 10-15$ GeV. At higher 
energies the UE $dN_{ch}/d\eta$ is flatten and $R_{f}$ and $T_{0}$
are very close to that given by (\ref{eq:60}) and (\ref{eq:70}).
And radiative and collisional energy loss may be calculated using
(\ref{eq:60}), (\ref{eq:70}). To illustrate it in Fig.~1 we presented
the results for the total energy loss obtained for the fireball
parameters for the UE $dN_{ch}/d\eta$ in the plateau region
($p_{T}^{l}\gsim 5$ GeV). From Fig.~1 one can see that
the energy loss for these two versions of the fireball
parameters (solid and long-dashed lines) become very close to each
other at $E\gsim 10$ GeV. This says that the decrease of the 
UE multiplicity density at $p_{T}^{l}\lsim 5$ GeV should be practically 
unimportant for $R_{pp}(p_{T})$ already at $p_{T}\gsim 7-10$ GeV.
\begin{figure}[h]
\begin{center}
\epsfig{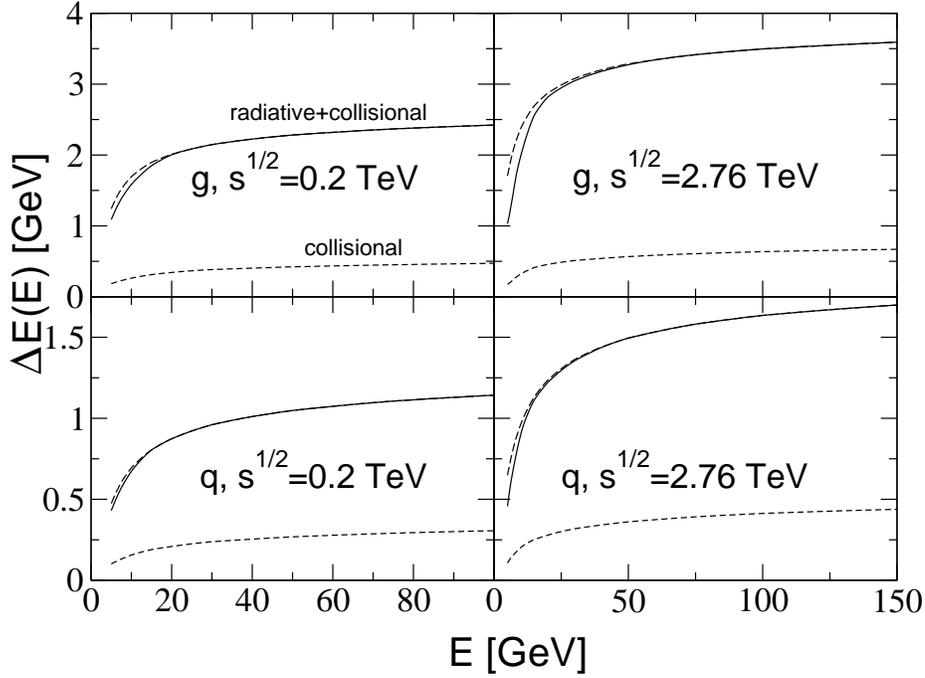}
\end{center}
\caption[.]
{
Energy dependence of the  
energy loss of gluons (upper panels) and light quarks (lower
panels) produced in the center of the mini-QGP 
fireball at $\sqrt{s}=0.2$ TeV (left) and $\sqrt{s}=2.76$ TeV (right). 
Solid line: total (radiative plus collisional) energy loss 
calculated  with the fireball radius $R_{f}$ and the initial temperature $T_{0}$
obtained with the UE $dN_{ch}/d\eta$ dependent on the initial parton
energy $E$; dashed line: same as solid line but for collisional energy loss;
 long-dashed line: same as solid line but for $R_{f}$ and $T_{0}$ 
obtained with the
UE $dN_{ch}/d\eta$ in the plateau region as given by (\ref{eq:60})
and (\ref{eq:70}).
All the curves are for $\alpha_{s}^{fr}=0.6$. 
}
\end{figure}
\begin{figure*}[t]
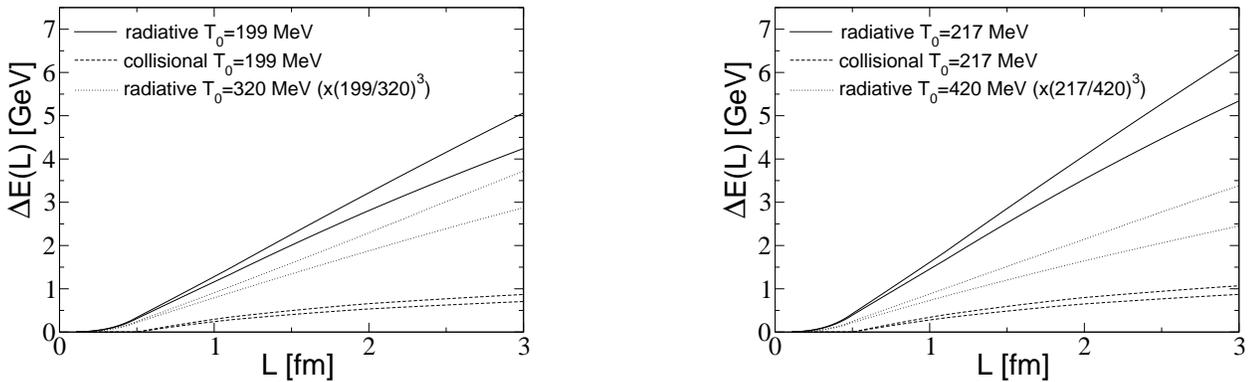

\hspace*{-0.8cm }\epsfig{file=fig2na.eps,height=5cm,clip=,angle=0} 
\hspace*{0.8cm } \epsfig{file=fig2nb.eps,height=5cm,clip=,angle=0} 
\begin{minipage}[t]{17.cm}  
\begin{center}
\caption{
Left:
Radiative (solid) and collisional (dashed) 
gluon energy loss   vs the path length $L$ in the 
QGP with $T_{0}=199$ MeV for (bottom to top)  $E=20$ and $50$ GeV.
The dotted lines show radiative energy loss for $T_{0}=320$ MeV
rescaled by the factor $(199/320)^{3}$. 
All curves are calculated for $\alpha_{s}^{fr}=0.6$.
Right: same as in the left figure but for $T_{0}=217$ and $420$ MeV
and the rescaling factor $(217/420)^{3}$ for dotted lines. 
}
\end{center}
\end{minipage}
\end{figure*}

Fig.~1 shows that the parton energy loss in the mini-QGP turns out to 
be quite large. At $E\sim 10-20$ GeV for gluons the total energy loss 
is $\sim 10-15$\% of the initial energy.
The contribution of the collisional mechanism is relatively small. 
The energy loss for the mini-QGP shown in Fig.~1 is smaller than
that obtained in \cite{RAA13} for the large-size QGP in $AA$ collisions
by only a factor of $\sim 4$. 

To illustrate 
the $L$-dependence of the parton energy loss in our model
in Fig.~2 we show the results 
for the radiative and collisional gluon energy loss vs 
the path length $L$
for $E=20$ and $50$ GeV
for $T_{0}=199$ and $217$ MeV, corresponding to $\sqrt{s}=0.2$ 
and $2.76$ TeV.
To show the difference 
between the QGP produced in $pp$ and $AA$ collisions 
we present also predictions for radiative energy loss 
for $T_{0}=320$ MeV corresponding to central $Au+Au$ collisions
at $\sqrt{s}=0.2$ TeV, and
for $T_{0}=420$ MeV corresponding to central $Pb+Pb$ collisions
at $\sqrt{s}=2.76$ TeV ( the procedure that leads to these values
of $T_{0}$ is described in \cite{RAA13}). 
To illustrate the temperature dependence
better we rescaled the predictions for $AA$ collisions by 
the factor $(T_{0}(pp)/T_{0}(AA))^{3}$.
One can see that at $L\ge \tau_{0}$ the radiative energy loss
is approximately a linear function of $L$. At $L<\tau_{0}$
the radiative energy loss is approximately $\propto L^{3}$
(since the leading $N=1$ rescattering contribution to the effective
Bethe-Heitler cross section is $\propto L$ \cite{Z_OA,AZ} and integration 
over the longitudinal coordinate of the scattering center gives additional 
two powers of $L$).
The comparison of the radiative energy loss for $T_{0}=199$ and $217$ 
MeV to that for $T_{0}=320$ and $420$ MeV shows deviation from the $T^{3}$ 
scaling by factors of $\sim 1.5$ and $\sim 2$, respectively. 
One can see that this difference persists even 
at $L\sim 1$ fm.
This deviation from the $T^{3}$ scaling comes mostly from
the increase of the LPM suppression (and partly from the increase
of the Debye mass) for the QGP produced in $AA$ collisions.

\subsection{Jet quenching in $pp$ collisions}
As we have seen from Fig.~1 the dependence of 
the UE multiplicity density from
\cite{PHENIX_dA,ATLAS_UE_Nch} on the momentum of the leading jet hadron $p_{T}^{l}$
practically does not affect  the parton energy loss 
at $E\gsim 10-15$ GeV, which from the standpoint of the particle spectra
corresponds approximately to $p_{T}\gsim 7-10$ GeV.
To account for the effect of the $p_{T}^{l}$-dependence of
the UE $dN_{ch}/d\eta$ on $R_{pp}(p_{T})$ at $p_{T}\lsim 10$ GeV we use the rigid approximation 
$p_{T}^{l}=\langle z_{l}\rangle E$ as in the above 
calculations of the energy loss. And in addition 
ignore the fluctuations of the variable $z$ in (\ref{eq:80})
(since the integrand of (\ref{eq:80}) is quite
sharply peaked about $\langle z \rangle$).
In this approximation we can write $p_{T}^{l}= p_{T}/\eta$, 
where $p_{T}$ is the momentum
of the observed particle in (\ref{eq:80}) and 
$\eta=\langle z\rangle/\langle z_{l} \rangle$. 
Jet simulation with PYTHIA \cite{PYTHIA} shows that for jets with 
energy $E\lsim 10-15$ GeV, that can feel the energy dependence 
of the UE multiplicity, one can take $\eta\sim 2.1$ for $\sqrt{s}=0.2$ TeV
and $\eta\sim 1.9$ at LHC energies $\sqrt{s}\gsim 2.76$ TeV. 
The uncertainties from this prescription
are restricted to the region $p_{T}\lsim 7-10$ GeV. However, even in this
region it should work on average (in the sense that the fluctuations
will just smear the $p_{T}$-dependence of the medium suppression
in this region).  This problem becomes
completely irrelevant for $R_{pp}$ at $p_{T}\gsim 10$ GeV.
\begin{figure}[h]
\begin{center}
\epsfig{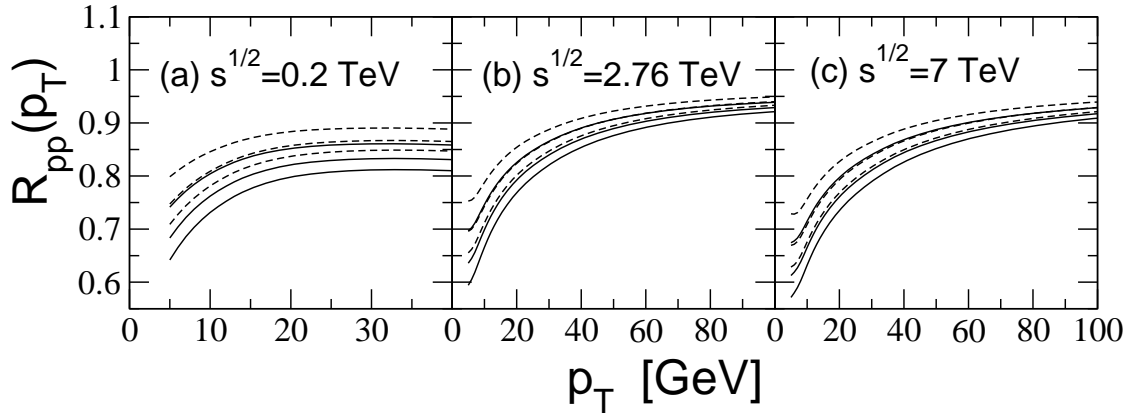}
\end{center}
\caption[.]
{$R_{pp}$ of charged hadrons 
at $\sqrt{s}=0.2$ (a), $2.76$ (b), $7$ (c) TeV for (top to bottom)
$\alpha_{s}^{fr}=0.5$, $0.6$ and $0.7$ for $\tau_{0}=0.5$ (solid)
and $0.8$ (dashed) fm.
}
\end{figure}

In Fig.~3 we present the results for $R_{pp}$ of charged hadrons 
at $\sqrt{s}=0.2$, $2.76$ and $7$ TeV for $\alpha_{s}^{fr}=0.5$, 
$0.6$ and $0.7$.
To illustrate the sensitivity of the results to $\tau_{0}$
we show the curves 
for $\tau_{0}=0.5$ and $0.8$ fm.
One can see that the suppression effect for 
the basic variant with $\tau_{0}=0.5$ fm
turns out to be quite large at $p_{T}\lsim 20$ GeV both for
RHIC and LHC. Fig.~3 shows that for $\tau_{0}=0.8$ fm the reduction
of the suppression is not very significant.
One can see that, as we expected, $R_{pp}$ does not exhibits a 
strong dependence on $\alpha_{s}^{fr}$. 
Although the plasma density is smaller at
$\sqrt{s}=0.2$ TeV,  the suppression effect is approximately similar 
to that at $\sqrt{s}=2.76$ and $7$ TeV. It is due to a steeper slope of the hard
cross sections at $\sqrt{s}=0.2$ TeV. The increase in the suppression
from $\sqrt{s}=2.76$ to $\sqrt{s}=7$ TeV is relatively small.

To understand the sensitivity of $R_{pp}$ to the fireball radius
we also performed the calculations for the fireball radii
calculated with (\ref{eq:50_1}), (\ref{eq:50}) times $0.7$ and $1.3$. 
We obtained in these two cases 
the reduction of the medium suppression by $\sim 3$\% and $10$\%, respectively.
The weak dependence on the value of $R_{f}$ is due to a compensation 
between the enhancement of the energy loss caused by increase of the 
fireball size and its suppression caused by reduction of the fireball density.
\footnote{The fact that for $0.7R_{f}$ and $1.3R_{f}$ the variations of $R_{pp}$
have the same sign is not very surprising since we use a wide 
window in the the fireball size. In this situation the second order 
term in the Taylor expansion of $R_{pp}$ around $R_{f}$ may be bigger 
than the linear term}. 
The stability of $R_{pp}$ against variations of $R_{f}$ gives a 
strong argument that the errors due to the neglect of the variation 
of the plasma density 
in the transverse coordinates should be small.
Indeed, the dominating $N=1$ rescattering contribution to the 
radiative energy loss is a linear functional of the 
plasma density profile along the fast parton trajectory. It means
that the results for a more realistic distribution
of the initial plasma density in
the impact parameter plane 
with a higher density in the central region can be roughly approximated 
by a linear superposition of the results obtained for the step 
density distributions (with different $R_{f}$) that should lead to approximately
the same $R_{pp}$ as our calculations.
Note also that since the variation of the plasma density in our test 
is very large (by a factor of $\sim 3.5$) this stability  at the same time
indicates indirectly that the effect of the neglected hydrodynamical 
variation of the plasma density should be small as well.

The results shown in Fig.~3 are obtained using the typical UE 
multiplicity density.
As we said in Sec.~3, an accurate accounting for the fluctuations
of the UE $dN_{ch}/d\eta$ is impossible since it should be done 
on the event-by-even basis (in the sense of the impact parameter 
and the jet production point), and
requires detailed information about dynamics of the 
UEs.
To understand how large the theoretical uncertainties, related to the
event-by-event fluctuations of the UE $dN_{ch}/d\eta$,
might be we evaluated $R_{pp}$ assuming 
that the distribution in the UE $dN_{ch}/d\eta$ is the same 
at each impact parameter and jet production point. We performed the
calculations using the distribution in $dN_{ch}/d\eta$ from CMS
\cite{CMS_UE_Nch} measured at $\sqrt{s}=0.9$ and $7$ TeV.
It obeys approximately KNO scaling low
similar to that in the minimum bias events \cite{Dumitru_KNO}.
For this reason one can expect that it can be used to estimate
the effect of the multiplicity fluctuations for RHIC conditions
as well. Our results show that for the fluctuating  $dN_{ch}/d\eta$
the magnitude of $(1-R_{pp})$ is reduced by only $\sim 5-6$\% 
both for RHIC and LHC energies. 
This says that our approximation without the event-by-event
fluctuations of the fireball parameters is quite good,
since it is very unlikely that an event-by-event analysis
may change significantly the results obtained using 
the total fluctuations of the UE multiplicity density.

\subsection{Jet quenching in $AA$ collisions}
Although $R_{pp}$ is unobservable quantity it can alter the results of 
the jet tomography of $AA$ collisions. 
To illustrate the possible effect of the mini-QGP
in $pp$ collisions on $R_{AA}$ we show in Fig.~4
the comparison of our results for $R_{AA}$
with the data for $\pi^{0}$-mesons in central $Au+Au$ 
collisions at $\sqrt{s}=0.2$ TeV (a) from PHENIX \cite{PHENIX_RAA_pi},
and with the data for charged hadrons in central $Pb+Pb$ collisions
at $\sqrt{s}=2.76$ TeV (b,c)
from ALICE \cite{ALICE_RAAch} and CMS \cite{CMS_RAAch}. 
\begin{figure} [t]
\begin{center}
\epsfig{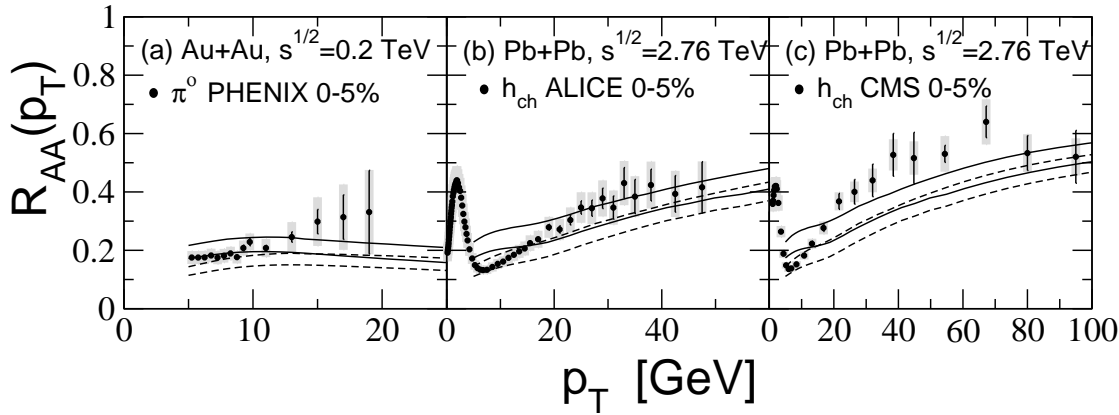}
\end{center}
\caption
{(a) $R_{AA}$ of $\pi^{0}$ for 0-5\% central $Au+Au$ collisions
at $\sqrt{s}=0.2$ TeV from our calculations 
for (top to bottom) $\alpha_{s}^{fr}=0.5$ and $0.6$
with (solid) and without (dashed) $1/R_{pp}$ factor in (\ref{eq:30}).
(b,c) $R_{AA}$ for charged hadrons for 0-5\% central $Pb+Pb$ collisions
at $\sqrt{s}=2.76$ TeV from our calculations 
for (top to bottom) $\alpha_{s}^{fr}=0.4$ and $0.5$
with (solid) and without (dashed) $1/R_{pp}$ factor in (\ref{eq:30}).
The solid curves are obtained with the factor $1/R_{pp}$ 
calculated with $\alpha_{s}^{fr}=0.6$.
Data points are from  PHENIX \cite{PHENIX_RAA_pi} (a),
ALICE \cite{ALICE_RAAch} (b) and CMS \cite{CMS_RAAch} (c).
Systematic experimental errors are shown as shaded areas. 
}
\end{figure}
We show the results obtained with (solid) the $1/R_{pp}$ factor,
i.e. for $R_{AA}$ defined by (\ref{eq:30}),
and the results without (dashed) this factor,
i.e. for $R_{AA}^{st}$.
We use the $R_{pp}$ obtained with $\alpha_{s}^{fr}=0.6$.
We present the curves for $R_{AA}^{st}$ obtained with
$\alpha_{s}^{fr}=0.5$ and $0.6$ for $\sqrt{s}=0.2$ TeV, 
and with $\alpha_{s}^{fr}=0.4$ and $0.5$
for $\sqrt{s}=2.76$ TeV. Since these values give  better agreement 
with the data of the $R_{AA}$ given by (\ref{eq:30}).
In calculating the hard cross sections for $AA$ collisions we account for 
the nuclear modification of the PDFs
with the EKS98 correction \cite{EKS98}.
As in \cite{RAA13} we take $T_{0}= 320$ MeV for central
$Au+Au$ collisions at $\sqrt{s}=0.2$ TeV, and
$T_{0}= 420$ MeV for central
$Pb+Pb$ collisions at $\sqrt{s}=2.76$ TeV
obtained from hadron multiplicity pseudorapidity density $dN_{ch}/d\eta$ 
from RHIC \cite{STAR_Nch} and LHC \cite{CMS_Nch,ALICE_Nch}.
One can see that at $p_{T}\sim 10$ GeV for RHIC the agreement 
of the theoretical $R_{AA}$
(with the  $1/R_{pp}$ factor) with the data  
is somewhat better for $\alpha_{s}^{fr}=0.6$, and for LHC 
the value $\alpha_{s}^{fr}=0.5$ seems to be preferred by the data.
But the agreement in the $p_{T}$-dependence
of $R_{AA}$ is evidently not perfect (especially for LHC).
One sees that the theory somewhat
underestimates the slope of the data. 
And the regions of large $p_{T}$ support 
$\alpha_{s}^{fr}=0.5$ and $0.4$ for RHIC and LHC, respectively.
One can see that the inclusion
of $R_{pp}$ even reduces  a little  the slope of $R_{AA}$
(since $R_{pp}$ in the denominator on the right hand side of
 (\ref{eq:30}) grows with $p_{T}$).
However, this discrepancy does not seem to be very dramatic
since the theoretical uncertainties 
may be significant.

From Fig.~4 one can see that the effect of $R_{pp}$ on $R_{AA}$ 
for the central $AA$ collisions can approximately be imitated by 
simple reduction of the $\alpha_{s}^{fr}$.
However, of course, it does not mean that all the theoretical predictions
for jet quenching in $AA$ collisions are insensitive to 
the medium modification of high-$p_{T}$ spectra in $pp$ collisions. 
It is clear that the effect of the $R_{pp}$ should be 
important for $v_{2}$ and centrality dependence of $R_{AA}$
(simply because in the scenario with the mini-QGP formation 
in $pp$ collisions the values of $\alpha_{s}^{fr}$ become bigger).
It should also be important for the flavor dependence of the 
the theoretical $R_{AA}$ since the suppression effect for heavy quarks 
in $pp$ collisions is smaller 
(by a factor of $\sim 1.5-2$  as our calculations show).
In the present exploratory study we do not consider these issues, and leave 
them for future work.

\subsection{Jet quenching in $pA$ collisions}
The medium suppression factor $R_{pp}$ should also be taken into
account in calculating the nuclear modification factor for $pA$ collisions.
Similarly to (\ref{eq:30}) the correct formula reads
 $R_{pA}=R_{pA}^{st}/R_{pp}$.
Comparison with data on $R_{pA}$ may be even more crucial for 
the scenario with the formation of the mini-QGP in $pp$ collisions 
since the sizes and the initial temperatures of the plasma fireballs in $pp$ and $pA$ collisions
should not differ strongly. And for this reason the predictions
for $R_{pA}$ should not have much uncertainties related to variation of
$\alpha_{s}$ or the temperature dependence of the plasma 
density and the Debye mass.
The ALICE data \cite{ALICE_RpPb}
on $R_{pPb}$ at $\sqrt{s}=5.02$ TeV exhibit a small deviation from 
unity at $p_{T}\gsim 10$ GeV, where the Cronin effect should be weak. 
In the scenario with the formation of the QGP in $pp$ and $pA$ collisions
this is possible only if the magnitudes of the medium suppression
in both the processes are very close to each other.
Unfortunately, presently we have not data on the UE multiplicity in 
$pPb$ collisions. However, it is clear that it cannot be smaller
than the minimum bias multiplicity density 
$dN_{ch}^{mb}/d\eta=16.81\pm 0.71$ \cite{ALICE_dnch_pPb}.
In principle, it is possible that in the typical minimum bias events
the energy deposited in the central rapidity region is 
already saturated due to a large number of 
the nucleons which interact with the proton in each $pPb$ collision,
and the enhancement of the multiplicity due to jet production 
is relatively small.
The preliminary PHENIX data \cite{PHENIX_dA} on the UE in 
$d Au$ collisions at $\sqrt{s}=0.2$ TeV
really indicate that for dominating small centralities
the enhancement of the  UE activity as compared to the minimum bias events 
is relatively small. 
In order to understand the restrictions on the
UE multiplicity density in $pPb$ collisions in the scenario with the mini-QGP
formation we simply calculate $R_{pPb}$ for 
$dN_{ch}/d\eta=K_{ue}dN_{ch}^{mb}/d\eta$ for 
$K_{ue}=1$, $1.25$, and $1.5$.
\begin{figure} []
\begin{center}
\epsfig{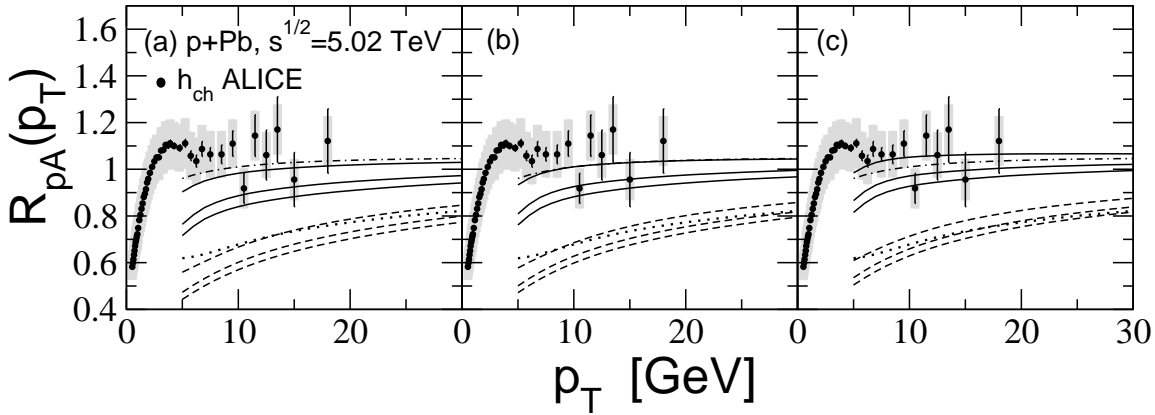}
\end{center}
\caption[.]
{(a) $R_{pPb}$ for charged hadrons at 
$\sqrt{s}=5.02$ TeV from our calculations for $\alpha_{s}^{fr}=0.6$
with (solid) and without (dashed) the $1/R_{pp}$ factor
for (top to bottom)  $K_{ue}=1$, $1.25$ and $1.5$ 
for the $R_{f}(pPb)$ from (\ref{eq:100}).  
(b,c) same as (a) but for $R_{f}(pPb)$ times $1.2$ and $1.4$.
The dotted line shows $R_{pp}$. The dot-dashed line
shows $R_{pPb}$  due to the EKS98 correction
\cite{EKS98} to the nucleus PDFs.
Data points are from ALICE \cite{ALICE_RpPb}. 
}
\end{figure}

To evaluate  $R_{pPb}$ we also need the fireball radius $R_{f}(pPb)$ 
which may be bigger than that in $pp$ collisions.
In our calculations as a basic choice we use the parametrization of the 
$R_{f}(pPb)$ as a function of the multiplicity density given in \cite{RPP}
obtained from the results of simulation of the $pPb$ collisions  
performed in \cite{glasma_pp}    
within the IP-Glasma model \cite{IPG12}.
The  $R_{f}(pPb)$ from \cite{IPG12} is close to the $R_{f}(pp)$
in the region where $R_{f}(pp)\propto(dN_{g}/dy)^{1/3}$, but flattens 
at higher values of the gluon density. Using the parametrization 
for $R_{f}(pPb)$ of Ref. \cite{RPP} and formula (\ref{eq:40}), we obtained
for our set of the enhancement factors
for the UE multiplicity $K_{ue}=[1,1.25,1.5]$
\beq
R_{f}(pPb)\approx[1.63,1.88,1.98]\,\,\mbox{fm}\,,
\label{eq:100}
\eeq
\beq
T_{0}(pPb)\approx[222,229,235]\,\,\mbox{MeV}\,.
\label{eq:110}
\eeq

Comparison of our results with the data on  
$R_{pPb}$ at $\sqrt{s}=5.02$ TeV
from ALICE \cite{ALICE_RpPb} is  shown in Fig.~5.
To illustrate the sensitivity to the fireball size in Fig.~5
we also present the results for the $R_{f}(pPb)$ $1.2$ and $1.4$ times
greater.
As for $AA$ collisions we show the curves with (solid) and without
(dashed) the $1/R_{pp}$ factor.
Similarly to $R_{AA}$ we account for the nuclear modification
of the PDFs with the EKS98 correction \cite{EKS98}
(which gives a small deviation of $R_{pPb}$ from unity even without
parton energy loss).
The results for $R_{pp}$ are also shown (dotted).
All the curves are obtained with $\alpha_{fr}=0.6$.
However, our predictions for $R_{pPb}$ (with the $1/R_{pp}$ factor)
are quite stable against variation of $\alpha_{s}^{fr}$ since
the medium suppression is very similar for $pp$ and $pPb$ collisions.

From Fig.~5 one can see that at $p_{T}\gsim 10$ GeV, where the 
Cronin effect should be small, our predictions (with $1/R_{pp}$ factor) 
obtained with $K_{ue}=1$ agree qualitatively with the data. The agreement 
becomes better with
increase of the $R_{f}(Pb)$. However, similarly to $R_{pp}$ the variation
of $R_{pPb}$ with the fireball size is relatively weak.
The curves for the higher UE multiplicities ($K_{ue}=1.25$ and $1.5$) lie below 
the data. 
Thus, Fig.~5 shows that the data from ALICE \cite{ALICE_RpPb}
may be consistent with the formation of the QGP in $pp$ and $pPb$ collisions
if the UE multiplicity is close to the minimum bias one.
This condition may be somewhat weakened if the size of the fireball in $pPb$ 
collisions is considerably bigger than predicted in \cite{glasma_pp}.
But it seems to be rather unrealistic since the required increase of 
the fireball size is too large. Say, for
a good agreement with the data for the UE multiplicity enhancement
factor $K_{ue}=1.5$ one should increase the $R_{f}(pPb)$ by a 
factor of $\sim 1.7$.
 
\subsection{A few remarks about approximations and robustness of the results}
One remark is in order about the description of the QGP in the ideal gas model
in our study. 
Of course, the QGP temperature formally defined in this model from
experimental multiplicity densities is somewhat incorrect.
Say, at $T\sim 200$ MeV the ideal gas formula for the 
entropy underestimates the plasma temperature by $10-15$\% 
as compared to the entropy from the lattice
calculations \cite{T_c1,T_c2}.
However, this fact is practically unimportant for our calculations.
Indeed, the potential $v(\rho)$ 
in the two-dimensional Schr\"odinger equation, 
which is used for evaluation of the induced gluon $x$-spectrum
in the LCPI approach \cite{LCPI}, is proportional to the entropy.
Since in our calculations in each case we use the 
entropy extracted directly from the experimental multiplicity densities, 
the temperature enters our calculations only through the Debye mass
in the dipole cross section. 
But the latter depends weakly on the Debye mass. For this
reason $10-15$\% errors in the temperature can be safely ignored.
An accurate definition of the temperature does not make much sense
since anyway the nonperturbative effects should modify the form
of the potential. 
Also, one should bear in mind that presently there are many other 
theoretical uncertainties in the jet quenching calculations,
and practically the theory cannot give absolute predictions for the 
medium suppression.
However, one can expect that it can be used to describe the variation of 
jet quenching from one experimental situation to another
(when the parameters of the model are already fitted to 
some experimental data). 
And we do follow this strategy in the present work. 
We calculate the medium suppression for the small-size
plasma in $pp$ collisions using the information about the values of 
$\alpha_{s}^{fr}$ which are necessary for description of the data on
$R_{AA}$. Without this information 
one could obtain only a very crude estimate of the effect.
Of course, the extrapolation from $AA$ to $pp$ collisions
assumes that for the real QGP the potential $v$ is 
approximately proportional to the entropy, as it is for the ideal QGP.
But this assumption seems to be physically very reasonable.
Anyway the extrapolation from $AA$ to $pp$ collisions 
should not give large errors since the difference of the plasma temperatures 
in these two cases is not very big.

It is worth to emphasize that for a reliable extrapolation of the theoretical
predictions from $AA$ to $pp$ collisions the calculations
should be performed with accurate treatment of the LPM suppression (which
is very important for $AA$ collisions) and finite-size and Coulomb effects 
(which are very important for the mini-QGP produced in $pp$ collisions).
Also, the calculations should be made with running $\alpha_{s}$ since the 
typical virtualities for induced gluon emission in the mini-QGP are  
higher than that in the large-size QGP in $AA$ collisions.
The LCPI \cite{LCPI} approach used in the present analysis satisfies all these
requirements.

Note that, in principle, 
the assumption that the produced QCD matter exists in the form of an
equilibrated QGP is not crucial
for our main result that there must
be a rather strong jet quenching in $pp$ collisions.
Since even if the created matter is some kind of 
a hadron resonance gas 
the parton energy loss will be approximately the same as for
the QGP because for a given entropy density the intensities
of multiple scattering for the hadron matter and the QGP
are very similar \cite{Baier_qhat}. 
Note that, since the most 
important quantity, which controls induced gluon emission,
is the number density of the color constituents in the medium,
from the standpoint 
of jet quenching, it is even not very important whether the
created QCD matter is equilibrated or not. 
Therefore one can say that in the pQCD picture of jet quenching
the significant medium suppression of hadron spectra in $pp$ collisions is 
an inevitable consequence of the observed UE multiplicities 
in $pp$ collisions and the medium suppression
of hadron spectra in $AA$ collisions (which allows to fix free parameters).

From the point of view of the pQCD the  medium
suppression of the high-$p_{T}$ spectra  in $pp$ collisions may 
be regarded as a higher twist effect. And of course it would be interesting
to observe it through a deviation of the experimental spectra 
from predictions of the standard pQCD formulas. But it is 
difficult since the medium suppression 
should have a very smooth onset in the energy region where 
the regime of free streaming hadrons transforms to a relatively 
slow collective expansion of the fireball.
Probably it could still occur
at $\sqrt{s}\sim 30-40$ GeV, where the UE 
pseudorapidity multiplicity density may be $\sim 2-3$ and $T_{0}\sim T_{c}$.
For this reason direct observation of 
this effect by comparing the pQCD predictions with experimental spectra 
is hardly possible since it is fairly hard to differentiate it from the
variations of the theoretical predictions related to
small modifications of the PDFs and of the FFs or other higher twist effects
not related to the mini-QGP. 
Also, presently the uncertainties of the pQCD predictions remain large 
\cite{Owens_NLO,Aurenche_NLO,Arleo_NLO1,Arleo_NLO2,Arleo_NLO3},
and the deviation of the ratio data/theory from unity 
at energies 
$\sqrt{s}\lsim 50$ GeV 
\cite{Owens_NLO,Aurenche_NLO,Arleo_NLO2} 
is often considerably bigger than the found medium effects.
In this situation it is difficult to identify a relatively small effect
from the mini-QGP. 
 Nevertheless, it worth noting that the results 
of the most recent NLO pQCD analysis of the inclusive charged particle
spectra in $pp$ and $\bar{p}p$ collisions at $\sqrt{s}=0.2-7$ TeV
\cite{NLO_Eskola} show that there is some deviation of the theory
from the data that seems to be qualitatively in line with the
the scenario with production of the mini-QGP which is more dense at LHC
energies. In \cite{NLO_Eskola} it was found that the LHC data prefer 
softer gluon FFs than the 
lower-$\sqrt{s}$ data.  But quantitatively the observed effect is 
considerably larger than what can be associated with the difference between
$R_{pp}$ at RHIC and LHC energies found in the present analysis.

It is worth noting that in principle the preliminary data from
ALICE \cite{ALICE_jet_UE} for $\sqrt{s}=7$ TeV 
support the existence of jet quenching in $pp$ collisions.
These data clearly indicate that the jet fragmentation becomes softer
with increase of the UE multiplicity. It is important that the effect is 
well seen already for the UE $dN_{ch}/d\eta$ smaller 
than the average one by a factor of $\sim 3$ 
(i.e., smaller than the typical UE $dN_{ch}/d\eta$ at $\sqrt{s}=0.2$ TeV).
Unfortunately, direct comparison of our calculations with the 
data \cite{ALICE_jet_UE}) is impossible, since there the NT90
jet observable has been used, which requires more detailed information on the
jet structure than our calculations can provide.

\section{Summary}
Assuming that a mini-QGP may be created in
$pp$ collisions, we have evaluated 
the medium modification factor $R_{pp}$ 
for light hadrons at RHIC ($\sqrt{s}=0.2$ TeV) and LHC 
($\sqrt{s}=2.76$ and  $7$  TeV) energies.
We have found an unexpectedly large suppression effect.
For $p_{T}\sim 10$ GeV we obtained
$R_{pp}\sim [0.7-0.8,\,0.65-0.75,\,0.6-0.7]$ 
at $\sqrt{s}=[0.2, 2.76,7]$ TeV.
We analyzed the role of the $R_{pp}$ in the
theoretical predictions for
the nuclear modification factor $R_{AA}$ in central $AA$ collisions at 
RHIC and LHC energies. We found that
the presence of $R_{pp}$ does not change dramatically 
the description of the data on  $R_{AA}$ for light hadrons
in central $AA$ collisions, and its effect may be imitated by some 
renormalization of $\alpha_{s}$.
Nevertheless, the effect of the QGP formation in $pp$ collisions 
may be potentially important in calculating other observables 
in $AA$ collisions. For example, it should affect 
$v_{2}$ and the centrality dependence of $R_{AA}$, 
and, due to the flavor dependence of $R_{pp}$,
its effect may be important for description of 
the flavor dependence of $R_{AA}$.
We leave analysis of these effects for future work. 
We also calculated the nuclear modification factor $R_{pPb}$ at
$\sqrt{s}=5.02$ TeV. Comparison with the data from ALICE \cite{ALICE_RpPb}
shows that the scenario with the formation of the QGP
in $pp$ and $pPb$ collisions may be consistent with the data
only if the UE multiplicity density in $pPb$ collisions
(which is unknown yet) is close to the minimum bias one.

\vspace {.7 cm}
\noindent
{\large\bf Acknowledgements}

\noindent
I am grateful to I.P.~Lokhtin for discussion of the results. I also
thank  D.V.~Perepelitsa
for useful information.
I am indebted to the referee, who pointed out on
the recent analysis \cite{NLO_Eskola}.
This work is supported 
in part by the 
grant RFBR
12-02-00063-a.


\section*{References}
\begin{thebibliography}{99}


\bibitem{GW}
M.~Gyulassy and X.N.~Wang,
Nucl. Phys. B{\bf 420}, 583 (1994).



\bibitem{BDMPS}
R.~Baier, Y.L.~Dokshitzer, A.H.~Mueller, S.~Peign\'e, and D.~Schiff,
Nucl.\ Phys.\ B{\bf 483}, 291 (1997); {\it ibid.} B{\bf 484}, 265 (1997).
%

\bibitem{LCPI}
B.G.~Zakharov, JETP\ Lett. {\bf 63}, 952 (1996); {\em ibid}
{\bf 65}, 615 (1997);
{\bf 70}, 176 (1999);
Phys.\ Atom.\ Nucl. {\bf 61}, 838 (1998).

\bibitem{BSZ}
R.~Baier, D.~Schiff, and B.G.~Zakharov, 
Ann.\ Rev.\ Nucl.\ Part. {\bf 50}, 37 (2000). 

\bibitem{W1}
U.A.~Wiedemann,
Nucl.\ Phys.\ A{\bf 690}, 731 (2001).


\bibitem{GLV1}
M.~Gyulassy, P.~L\'evai, and I.~Vitev, 
Nucl.\ Phys. B{\bf 594}, 371 (2001).

\bibitem{AMY}
P.~Arnold, G.D.~Moore, and L.G.~Yaffe,
JHEP {\bf 0206}, 030 (2002).





\bibitem{Bjorken1} J.D.~Bjorken, Fermilab preprint 
82/59-THY (1982, unpublished).

\bibitem{Arleo}
A.~Accardi, F.~Arleo, W.K.~Brooks, D.~D'Enterria, and 
V.~Muccifora,
 Riv.~Nuovo Cim. {\bf 32}, 439 (2010)
[arXiv:0907.3534].

\bibitem{Milhano}
Y.~Mehtar-Tani, J.G.~Milhano, and K.~Tywoniuk,
Int.~J.~Mod.~Phys. A{\bf 28,} 1340013 (2013). 

\bibitem{JW}
P.~Jacobs and  X.-N.~Wang,
Prog.~Part.~Nucl.~Phys. {\bf 54}, 443 (2005).


\bibitem{Armesto_HQ}
N.~Armesto,M.~Cacciari,A.~Dainese, C.A.~Salgado, and U.A.~Wiedemann,
Phys.~Lett. B{\bf 637}, 362 (2006).

\bibitem{BG} 	
A.~Buzzatti and  M.~Gyulassy,
Phys.~Rev.~Lett. {\bf108}, 022301 (2012). 

\bibitem{RAA12}
B.G.~Zakharov, 
JETP Lett. {\bf 96}, 616 (2013). 

\bibitem{RAA13}
B.G. Zakharov,
J. Phys. G{\bf 40}, 085003  (2013).





\bibitem{Bozek_pp} 	
P.~Bozek,
Acta Phys. Polon. B{\bf 41}, 837 (2010).

\bibitem{Wied_pp}	
J.~Casalderrey-Solana and U.A.~Wiedemann,
Phys. Rev. Lett. {\bf 104}, 102301 (2010). 


\bibitem{Werner}
K.~Werner, I.~Karpenko, T.~Pierog, M.~Bleicher, and K.~Mikhailov,
Phys.~Rev. C{\bf 83}, 044915  (2011). 



\bibitem{Camp1} 	
R.~Campanini, G.~Ferri, and G.~Ferri,
Phys. Lett. B{\bf 703}, 237 (2011).

\bibitem{Gyulassy_pp}
V.~Topor Pop, M.~Gyulassy, J.~Barrette, C.~Gale, and A.~Warburton,
Phys.~Rev. C{\bf 86}, 044902  (2012). 


\bibitem{glasma_pp}
A.~Bzdak, B.~Schenke, P.~Tribedy, and R.~Venugopalan,
arXiv:1304.3403.

\bibitem{SZ}
E.~Shuryak and I.~Zahed, arXiv:1301.4470.

\bibitem{CMS_ridge1}
V.~Khachatryan {\sl et al.}
[CMS Collaboration],
JHEP {\bf 1009}, 091 (2010). 

\bibitem{VH}
L. Van Hove, Phys. Lett. B{\bf 118}, 138 (1982).

\bibitem{STAR_fem} 	
M.M.~Aggarwal {\it et al.} [STAR Collaboration],
Phys.~Rev. C{\bf 83}, 064905 (2011).

\bibitem{ALICE_fem}
K.~Aamodt {\it et al.} [ALICE Collaboration],
Phys.~Rev. D{\bf 84}, 112004  (2011).

\bibitem{ALICE_jet_UE} 	
H.L.~Vargas,  for the 
ALICE Collaboration,
J. Phys. Conf. Ser. {\bf 389}, 012004  (2012) 
[arXiv:1208.0940].



\bibitem{Field}
R.~Field,
Acta Phys.~Polon. B{\bf 42}, 2631  (2011). 


\bibitem{CDF}
A.A. Affolder {\it et al.} [CDF Collaboration], Phys. Rev. D{\bf 65}, 092002
(2002).


\bibitem{Z_pp}
B.G. Zakharov, Phys.~Rev.~Lett. {\bf 112}, 032301 (2014). 

\bibitem{ALICE_RpPb}
B.~Abelev {\it et al.} [ALICE Collaboration],
Phys. Rev. Lett. {\bf 110}, 082302  (2013). 


\bibitem{RAA08}
B.G.~Zakharov, JETP Lett. {\bf 88}, 781 (2008). 

\bibitem{RAA11}
B.G.~Zakharov,
JETP Lett. {\bf 93}, 683 (2011).

\bibitem{Bjorken2}
J.D.~Bjorken, 
Phys.\ Rev. D{\bf 27}, 140 (1983).


\bibitem{Heinz_hydro} 	
H.~Song, S.A.~Bass, U.~Heinz, T.~Hirano, C.~Shen,
Phys.~Rev. C{\bf 83}, 054910 (2011), Erratum-ibid. C{\bf 86}, 059903 (2012). 


\bibitem{BM-entropy}
B.~M\"uller and K.~Rajagopal,
Eur. Phys. J. C{\bf 43}, 15 (2005).



\bibitem{IPG12}
B.~Schenke, P.~Tribedy, and R.~Venugopalan,
Phys. Rev. Lett. {\bf 108}, 252301 (2012). 

\bibitem{RPP} 	
L. McLerran, M.~Praszalowicz, and B.~Schenke,
arXiv:1306.2350.



\bibitem{PHENIX_dA}
J.~Jia, for the PHENIX Collaboration, contribution to the Quark Matter
2009 Conf., March 30 - April 4, Knoxville, Tennessee;
 arXiv:0906.3776.


\bibitem{ATLAS_UE_Nch}
G.~Aad {\it et al.}
[ATLAS Collaboration], 
Phys. Rev. D{\bf 83},  112001 (2011).




\bibitem{CMS_UE_Nch}
S.~Chatrchyan {\it et al.}
[CMS Collaboration],
JHEP {\bf 1109}, 109 (2011).

\bibitem{ALICE_UE_Nch}
B.~Abelev {\it et al.}  [ALICE Collaboration]
JHEP {\bf 1207}, 116 (2012). 







\bibitem{STAR-dnch}
B.I.~Abelev {\it et al.}
[STAR Collaboration], 
Phys.~Rev. C{\bf 79}, 034909 (2009).



\bibitem{T_c1}
P.~Petreczky,
J.~Phys. G{\bf 39}, 093002  (2012). 

\bibitem{T_c2}
S.~Borsanyi,
Nucl.~Phys.A{\bf 904-905}, 270c (2013) 
[arXiv:1210.6901].


\bibitem{BMS_hydro}
R.~Baier, A.H.~Mueller, and D.~Schiff,
Phys. Lett. B{\bf 649}, 147 (2007).



\bibitem{CTEQ6}
S.~Kretzer, H.L.~Lai, F.~Olness, and W.K.~Tung,
Phys. Rev. D{\bf 69}, 114005 (2004).




\bibitem{PYTHIA}
T.~Sjostrand, L.~Lonnblad, S.~Mrenna, and  P.~Skands,
arXiv:hep-ph/0308153.

\bibitem{KKP}
B.A.~Kniehl, G.~Kramer, and B.~Potter, 
Nucl.\ Phys. B{\bf 582}, 514 (2000).

\bibitem{BDMS_RAA}
R.~Baier, Yu.L.~Dokshitzer, A.H.~Mueller, and
D.~Schiff, JHEP {\bf 0109}, 033 (2001). 



\bibitem{Z_Ecoll}
B.G.~Zakharov,
JETP Lett. {\bf 86}, 444 (2007).

\bibitem{AZ_glasma}
P.~Aurenche and B.G.~Zakharov,
Phys.~Lett. B{\bf 718}, 937 (2013).




\bibitem{Bielefeld_Md}
O.~Kaczmarek and F.~Zantow,
Phys. Rev. D{\bf 71}, 114510 (2005).

\bibitem{LH}
P.~L\'evai and U.~Heinz,
Phys.\ Rev.\ C{\bf 57}, 1879 (1998).

\bibitem{Z04_RAA}
B.G.~Zakharov, JETP Lett. {\bf 80}, 617 (2004).

\bibitem{Z_OA}
B.G.~Zakharov, JETP Lett. {\bf 73}, 49 (2001).


\bibitem{NZ_piki}
N.N.~Nikolaev and B.G.~Zakharov,
Phys.~Lett. B{\bf 332}, 184 (1994).



\bibitem{NZ_HERA}
N.N.~Nikolaev and B.G.~Zakharov,
Phys. Lett. B{\bf 327}, 149 (1994). 

\bibitem{DKT}
Yu.L.~Dokshitzer, V.A.~Khoze, and S.I.~Troyan,
Phys.\ Rev. D{\bf 53}, 89 (1996).

\bibitem{AZ}
P. Aurenche and B.G. Zakharov, 
JETP Lett. {\bf 90}, 237 (2009)
[arXiv:0907.1918].




\bibitem{Dumitru_KNO}
A.~Dumitru and E.~Petreska, arXiv:1209.4105.


\bibitem{PHENIX_RAA_pi}
%
A. Adare {\it et al.}
[PHENIX Collaboration], arXiv:1208.2254.

\bibitem{ALICE_RAAch}
B. Abelev {\it et al.} [ALICE Collaboration],
Phys. Lett. B{\bf 720}, 52 (2013).



\bibitem{CMS_RAAch}
S. Chatrchyan {\it et al.} [CMS Collaboration],
Eur. Phys. J. C{\bf 72},  1945 (2012). 



\bibitem{EKS98}
K.J.~Eskola, V.J.~Kolhinen, and C.A.~Salgado,
Eur. Phys. J. C{\bf 9}, 61 (1999).

\bibitem{STAR_Nch}
B.I. Abelev {\it et al.}
[STAR Collaboration ],
Phys. Rev. C{\bf 79}, 034909 (2009).

\bibitem{CMS_Nch}
S. Chatrchyan {\it et al.} [CMS Collaboration], 
JHEP {\bf 1108}, 141 (2011).


\bibitem{ALICE_Nch}
K.~Aamodt {\it et al.} [ALICE Collaboration],
Phys. Rev. Lett. {\bf 106}, 032301 (2011).




\bibitem{ALICE_dnch_pPb}
B.~Abelev {\it et al.} [ALICE Collaboration],
Phys.~Rev.~Lett. {\bf 110}, 032301  (2013).




\bibitem{Baier_qhat}
R.~Baier,
Nucl.~Phys. A{\bf 715}, 209 (2003).


\bibitem{Owens_NLO} 	
L.~Apanasevich {\it et al.},
Phys.~Rev. D{\bf 59}, 074007  (1999).

\bibitem{Aurenche_NLO} 	
P.~Aurenche, M.~Fontannaz, J.P.~Guillet, B.A.~Kniehl, and M.~Werlen,
Eur.~Phys.~J. C{\bf 13}, 347 (2000).

\bibitem{Arleo_NLO1}
F.~Arleo,
JHEP {\bf 0609}, 015 (2006).

\bibitem{Arleo_NLO2}
F.~Arleo and D. d'Enterria,
Phys.~Rev. D{\bf 78}, 094004  (2008).

\bibitem{Arleo_NLO3}
F.~Arleo, D. d'Enterria, and A.S.~Yoon,
JHEP {\bf 1006}, 035 (2010).

\bibitem{NLO_Eskola}
D. d'Enterria, K.J. Eskola, I. Helenius, and H. Paukkunen ,
arXiv:1311.1415.

\end{thebibliography}
\end{document}